# Global Characterization of a Laser-Generated Neutron Source

D. P. Higginson[1,2], R. Lelièvre[1,3], L. Vassura[1,4], M. M. Gugiu[5], M. Borghesi[6], L.A. Bernstein[2,7,8], D. L. Bleuel[2], B. L. Goldblum[7,8], A. Green[6], F. Hannachi[9], S. Kar[6], S. Kisyov[2,5], L. Quentin[1], M. Schroer[10], M. Tarisien[9], O. Willi[10], P. Antici[4,11,12], F. Negoita[5], A. Allaoua[3] and J. Fuchs[1].

1) Laboratoire pour l'Utilisation des Lasers Intenses, UMR 7605 CNRS-CEA-École Polytechnique-Université Paris VI, 91128 Palaiseau, France
2) Lawrence Livermore National Laboratory, Livermore, California 94551, USA
3) Laboratoire de micro-irradiation, de métrologie et de dosimétrie des neutrons, PSE-Santé/SDOS, IRSN, 13115 Saint-Paul-Lez-Durance, France
4) Dipartimento SBAI, Università di Roma "La Sapienza," Via Scarpa 14-16, 00161 Roma, Italy
5) Institute of Atomic Physics, IFIN-HH, Bucharest-Magurele, P.O. Box MG6, Romania
6) The Queen's University of Belfast, Belfast BT7 1NN, United Kingdom
7) Department of Nuclear Engineering, University of California, Berkeley, California 94720, USA
8) Lawrence Berkeley National Laboratory, Berkeley, California 94720, USA
9) Centre d'Études Nucléaires de Bordeaux Gradignan, Université Bordeaux1, CNRS-IN2P3 Route du solarium, 33175 Gradignan, France
10) Institut für Laser und Plasmaphysik, Heinrich Heine Universität Düsseldorf, D-40225 Düsseldorf, Germany
11) INRS-EMT, Varennes, Québec, Canada
12) Istituto Nazionale di Fisica Nucleare, Via E. Fermi, 40-00044 Frascati, Italy

**Abstract**

Laser-driven neutron sources are routinely produced by the interaction of laser-accelerated protons with a converter. They present complementary characteristics to those of conventional accelerator-based neutron sources (e.g. short pulse durations, enabling novel applications like radiography). We present here results from an experiment aimed at performing a global characterization of the neutrons produced using the Titan laser at the Jupiter Laser Facility (Livermore, USA), where protons were accelerated from 23µm thick plastic targets and directed onto a LiF converter to produce neutrons. For this purpose, several diagnostics were used to measure these neutron emissions, such as CR-39, activation foils, Time-of-Flight detectors and direct measurement of $^7$Be residual activity in the LiF converters. The use of these different, independently operating diagnostics enables comparison of the various measurements performed to provide a robust characterization. These measurements led to a neutron yield of $2.0 \times 10^9$ neutrons per shot with a modest angular dependence, close to that simulated.

## I. INTRODUCTION

Ultra-intense lasers [1] have been used for more than two decades to produce neutrons [2,3] with less radiological constraints than conventional neutron sources, such as reactors or accelerator-based spallation sources. Indeed, the former produce large quantities of radioactive waste due to the fission reactions and the significant activation of their constituent materials [4,5]. Laser-driven neutron sources are characterized by significant emission of neutrons, from sub-MeV to tens or hundreds of MeV, in small time intervals ($<$ ns) leading to high-brightness short neutron pulses [6,7,8,9,10]. These features are of potential interest for many applications in astrophysics [11,12], plasma physics [13,14], medical sciences [15,16], security [17,18], industry [19] or non-destructive material analysis [20,21].

Several mechanisms can be used to produce neutrons using lasers, among them: photoneutron generation [22], beam-fusion reactions [23,24] or the pitcher-catcher technique. The principle of the pitcher-catcher technique has been widely demonstrated [25,26,27,28,29]. It allows for the generation of neutrons via the interaction of laser-accelerated ions (usually protons) with a second target called the converter. The selection of this converter depends on the proton energy distribution. A low-Z material (like Li or Be) should be selected if the proton spectrum is peaked at energies of a few MeV, to take advantage of the cross-sections of (p,n) reactions at these energies. However, a high-Z material is preferred if the proton spectrum has a larger component at energies of tens of MeV, so as to benefit from spallation reactions [30].

The laser-acceleration of ions is itself induced by laser-generated hot electrons that form a dense sheath on the solid target irradiated by the laser; this is the so-called Target Normal Sheath Acceleration (TNSA) mechanism [31,32,33,34].

We present here the results of an experiment carried out at the Titan Laser and aimed at performing a global characterization of laser-driven neutrons. The neutrons were produced by the interaction of protons, accelerated from plastic targets, with a LiF converter. The neutron emissions were characterized using a set of diagnostics, including organic plastic scintillators used as neutron Time-of-Flight (nToF) detectors, CR-39 and activation foils, and also simulated with Monte-Carlo transport codes MCNP6 [35] and Geant4 [36].

Section II gives an overview of the experimental setup and the characteristics of the proton spectrum used to produce neutrons. Section III describes how the neutron emissions were simulated using a Monte-Carlo code. Section IV shows the experimental measurements obtained using the different diagnostics and the comparisons with the simulated results. Finally, Section V summarizes the main results and concludes this study.

## II. EXPERIMENTAL OVERVIEW

### A. Experimental setup

The experiment was performed using the Titan Laser at the Jupiter Laser Facility at Lawrence Livermore National Laboratory, delivering pulses with 1054 nm wavelength and 650 fs duration. The laser was focused using an f/3 off-axis parabola to achieve an 8 μm full-width-at-half-maximum (FWHM) focal spot. With an on-target energy of ∼100 J, the on-target peak intensity was around $10^{20}$ W/cm$^2$.

As shown in Figure 1, the beam was focused onto a 23 μm polyethylene terephthalate (PET) target that was aluminized with a few microns on the target surface to increase laser absorption. The laser accelerated protons emanated from the rear surface of the target using the TNSA mechanism. These protons were aimed into a 2 mm thick LiF slab placed 20 mm away, where they caused the generation of neutrons via nuclear reactions:

$^7$Li(p,n)$^7$Be, Q = -1.644 MeV, threshold = 1.880 MeV
$^{19}$F(p,n)$^{19}$Ne, Q = -4.021 MeV, threshold = 4.235 MeV

The neutrons were characterized with a variety of diagnostics to measure their yield, such as CR-39, neutron activation of various isotopes (i.e. $^{115}$In, $^{27}$Al, $^{56}$Fe) and a direct measurement of the neutron generation by measuring the amount of residual Be activity produced in the LiF slab.

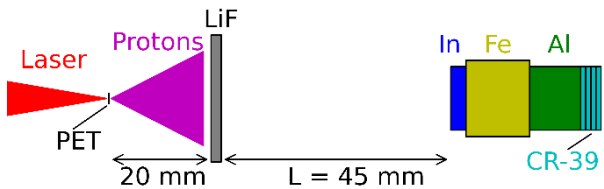

**Fig. 1.** Scheme of the diagnostic setup for the activation measurements, where the activation stack was placed at its closest distance to the LiF (shot 42). In the other setup, the stack was placed further back such that L = 205mm (shot 25).

The energy and angular dependence of the neutron yields were also experimentally studied using organic plastic scintillators (BC-400) for nToF measurements. As shown in Figure 2, a number of BC-400 scintillators with dimensions of 4x4x12 cm$^3$ were placed at different angles and distances between 2 and 6 meters relative to the proton beam direction and secondary LiF target, respectively. The photomultiplier tube used was Photonis XP2972 and its outputs were digitized for a duration of 1 μs at 1GS/s rate using the V1743 multichannel 12 bits switch-capacitor digitizer from CAEN.

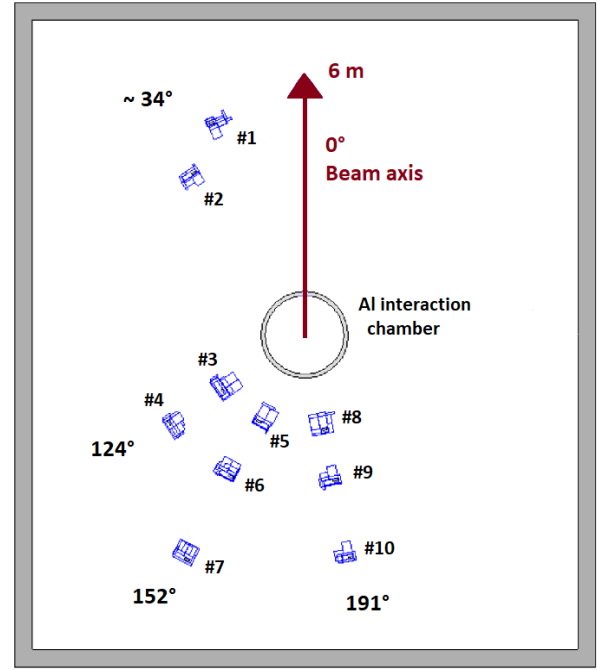

**Fig. 2.** Arrangement of the nToF detectors and their configurations of Pb shielding in the experimental hall.

### B. Proton spectrum

The proton spectrum was measured using layers of Gafchromic HD Radiochromic Film (RCF) [37] separated by filters of PET placed at 35 mm behind the target, in place of the LiF converter. The dose deposition was modelled using stopping powers taken from SRIM [38]. As shown in Figure 3, the proton spectrum, $\frac{dN_p}{dE}(E_p)$, was well-fit using an exponential spectrum:

$$\frac{dN_p}{dE}(E_p \leq E_{max}) = \frac{N_0}{T} \exp\left(-\frac{E}{T}\right) \quad (1)$$

$$\frac{dN_p}{dE}(E_p > E_{max}) = 0 \quad (2)$$

where $E_p$ is the proton energy, $N_0 = 2.14 \times 10^{12}$ is the total number of protons, $T = 6.3$ MeV is the slope temperature and $E_{max} = 30.5$ MeV is the cut-off or maximum proton energy.

However, the typical TNSA proton spectra are Maxwellian distributions, usually described as a sum of two exponential functions with a slope at low energy greater than the one at high energy. Thus, considering a simple exponential function to describe the proton spectrum tends to underestimate the number of low-energy protons.

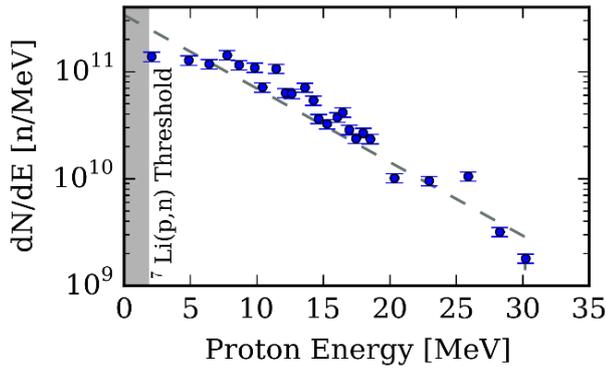

**Fig. 3.** Measured proton energy spectrum. The circles show the proton spectrum inferred using RCF and the dotted line is a decaying exponential, with parameters $N_0 = 2.14 \times 10^{12}$, $T = 6.3$ MeV and $E_{max} = 30.5$ MeV.

## III. MONTE-CARLO MODELING OF NEUTRON PRODUCTION

To simulate the production of neutrons in the LiF slab, the Monte-Carlo code Geant4 was used. This code includes particle scattering, energy loss and nuclear reactions. The proton-induced nuclear reactions included were $^7$Li(p,n)$^7$Be, $^6$Li(p,n)$^6$Be and $^{19}$F(p,n)$^{19}$Ne, with cross-sections taken from the ENDF/B-VIII.O library [39], which contains information on the angular dependence of these cross-sections. For example, the $^7$Li(p,n)$^7$Be cross section obtained from ENDF/B-VIII.O is shown in Figure 4.

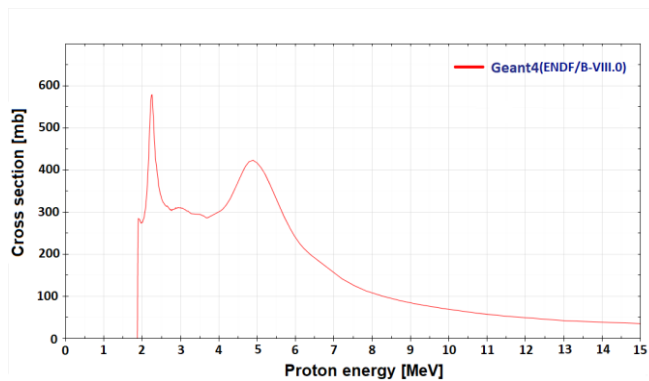

**Fig. 4.** Cross-section used for the $^7$Li(p,n)$^7$Be reaction.

As shown in Figure 5, the protons were injected in the simulation into a LiF target at a distance of 1 mm and allowed to propagate through a 2 mm thick LiF target (as in the experiment). The protons were given an exponential energy spectrum as described in Section II B. They were injected at the center of the LiF, which is a simplification because the proton beam should have a diameter of approximately 20 mm in the experiment due the divergence of the beam. The protons were injected with straight trajectories, which is another simplification as the proton beam will have a divergence angle. Note that, the measurement of the proton spectrum shown in Figure 3 and the neutron conversion in the LiF converter were performed in different shots. Due to the shot-to-shot variability of the laser parameters, the real proton spectrum injected in the converter to produce the neutrons could differ from the one provided in Figure 3.

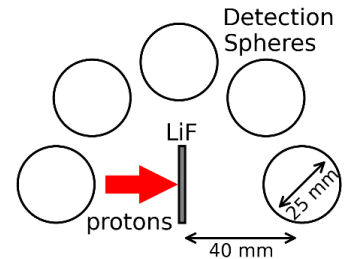

**Fig. 5.** Schematic of the setup used for the Geant4 simulations. The arrow shows the direction of the proton injection. The detection spheres are shown as circles.

Spherical detectors were placed in the simulations at angles of 0°, 45°, 90°, 135° and 180°, as shown in Figure 5. These detectors were approximately 40 mm away from the front surface of the LiF and had diameters of 25 mm. Neutrons passing through these detectors were binned in energy and normalized by the solid angle of the detectors.

Figure 6 shows the simulated yields of the detectors resolved in angle. For the injected proton spectrum, the neutron yield is relatively flat as a function of angle, as varying only by about a factor of 2.5. Inclusion of the divergence angle of the protons would result in a more isotropic distribution.

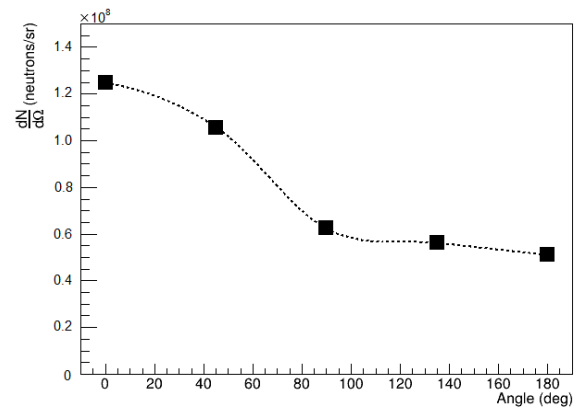

**Fig. 6.** The neutron yields found from Geant4 simulations.

Figure 7 shows the energy spectra of the neutrons in the different detectors. The peak in the cross-section which starts at around 4 MeV at 0° moves to lower energies as the angle increases, which is due to the kinematics of the interaction. Also, the higher energy neutrons are more likely to be observed at forward-directed angles. Again, including the proton angular divergence would minimize these effects.

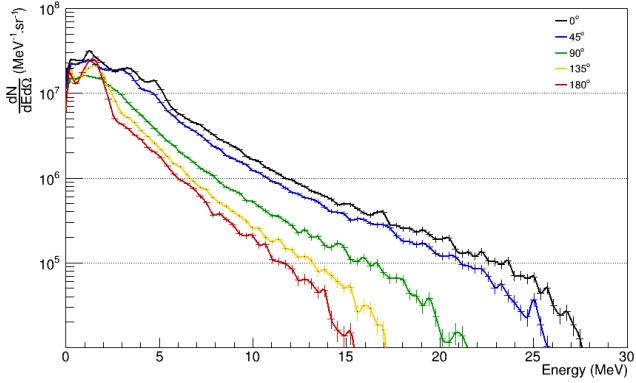

**Fig. 7.** The neutron spectrum at different angles as found from the Geant4 simulations.

These neutron spectra will next be used to model the responses of the diagnostics. The total number of neutrons recorded in these simulations was $9.80 \times 10^8$ neutrons. Additional simulations were run to obtain the number of neutrons produced only from Li. This number was found to be $8.92 \times 10^8$, which can be compared to the activation found via measurement of the $^7$Be decay as shown in Table I.

## IV. Neutron yields diagnostics

### A. Activation measurements

#### 1. Setup and characteristics of activation measurements

The activation samples were counted at two locations using high purity Ge detectors. The first location was onsite at the Titan laser, and the second was at the LLNL's Nuclear Counting Facility (NCF). The NCF is a low-background facility located underground, beneath six feet of magnetite, housing a dozen high-purity germanium detectors each surrounded by 10 cm of pre-WWII lead to minimize contributions from environmental radiation. A customized program, GAMANAL, performs background subtractions and peak fits, correcting for individual detector efficiency, sample geometry and self-shielding of gamma-rays in the samples [40]. The detectors are regularly calibrated against NIST-traceable multi-energy point sources and large-area distributed sources. Automated sample positioning systems ensure accurate and repeatable counting geometries. At the onsite location, Ge detector absolute efficiencies were calculated to take into account the various geometries of the irradiated samples.

Gamma decays at 844 keV, 847 keV and 336 keV were used to measure neutron activation yields in samples of natural Al, natural Fe and $^{115}$In, respectively. The number of neutrons produced in the $^7$Li(p,n) reaction were measured using the $^7$Li 478 keV gamma line populated in the electron capture decay of the $^7$Be ground state [41]. At the end of each laser shot, the total number of radioactive nuclei present in the samples, were deduced from the number of decays measured during time intervals starting at the time after the laser shot. The branching ratios of each transition were taken into account.

#### 2. Neutron characterization via residuals in LiF

The direct measurement of the total neutron production is the measurement of the residual isotopes in the LiF (i.e. the product) following irradiation. The nuclear reactions that generate neutrons from the proton reactions are: $^7$Li(p,n)$^7$Be, $^6$Li(p,n)$^6$Be, and $^{19}$F(p,n)$^{19}$Ne, as shown in the top of Table II. Unfortunately, the half-lives of both $^6$Be and $^{19}$Ne are under 20 seconds and thus too fast to make measurements feasible ex-situ due the time (~ 10 minutes) needed to vent the chamber, though such measurements might be possible via in-situ counting in the future. However, the $^7$Be residual has a 53.2-days half-life, which is easily measured outside of the chamber. The measured activities are given in Table I.

To estimate the number of residual isotopes and their activity, we used Monte-Carlo simulations (detailed in Section III), which include the $^7$Li(p,n)$^7$Be reaction. These yield a total number of residuals of $8.92 \times 10^8$ and an activity of 134 Bq. Table I shows the activation measured from the decay of $^7$Be residuals. The $^7$Be decay was measured both onsite and at the NCF on shot 42. The latter provided a slightly lower yield relative to that measured using the onsite detectors by a factor of 0.93. Given the high fidelity of NCF, this was considered a correction factor for the onsite data. Shot 25 only had onsite data and was corrected by this factor.

|  | Activity [Bq] | N [particles] |
|---|---|---|
| Shot 25 | 247 | $1.64 \times 10^9$ |
| Shot 42 | 291 | $1.93 \times 10^9$ |

**Table I.** Experimental measurement of activity and number, N, of $^7$Be nuclei residuals from the $^7$Li(p,n)$^7$Be in the LiF slab.

From Table II, we observe that our simulations predict a total number of neutrons by, on average, a factor of 2 times lower than the neutrons observed experimentally. The origin of this difference could lie in the shot-to-shot variation of the proton spectrum, as discussed in Section III.

#### 3. Neutron characterization via activation samples

Additionally, a stack of activation foils with multiple materials (see Figure 1) was used to infer the neutron yields and, due to the energy dependence of the cross-sections, some energy resolution [42].

For this experiment, we investigated multiple neutron activation materials, as shown in Table II, to determine the best ones to use for this particular experiment. For this evaluation, we use the neutron spectrum calculated in Section III, and estimated the reactions, using the Monte-Carlo code MCNP, in a cylindrical puck of 25 mm diameter, 10 mm thickness placed 40 mm away from the LiF.

These were not the final sizes used in the experiment (i.e. they are different than Figure 1), but they help to compare the different materials and are consistent with the geometries required to fit inside the chamber and allow for measurements at multiple angles. The results of these estimations are shown in Table II.

| Initial | Abund. [%] | Reaction | Threshold [MeV] | Final | $T_{1/2}$ | γ-Energy [keV] | $N_{est}$ | $A_{est}(0)$ [Bq] | $A_{est}$(20 min.) [Bq] |
|---|---|---|---|---|---|---|---|---|---|
| $^{19}$F | 100 | (p,n) | 4.2 | $^{19}$Ne | 17.2 s | | $1\times10^8$ | $4\times10^6$ | $10^{-14}$ |
| $^6$Li | 7.5 | (p,n) | 5.9 | $^6$Be | < | | | | |
| $^7$Li | 92.5 | (p,n) | 1.9 | $^7$Be | 53.2 d | 478 | $1\times10^9$ | 150 | 150 |
| $^{115}$In | 95.7 | (n,n') | 0.34 | $^{115m}$In | 4.49 h | 336 | $2\times10^6$ | 85.8 | 81.5 |
| $^{27}$Al | 100 | (n,p) | 1.9 | $^{27}$Mg | 9.46 min. | 844 | $1\times10^4$ | 12.2 | 2.8 |
| $^{56}$Fe | 91.72 | (n,p) | 3.0 | $^{56}$Mn | 2.58 h | 847 | $6\times10^4$ | 4.5 | 4.1 |
| $^{58}$Ni | 68.08 | (n,p) | 0 | $^{58}$Co | 70.9 d | 811 | $2\times10^6$ | 0.23 | 0.23 |
| $^{58}$Ni | 68.08 | (n,d) | 6.1 | $^{57}$Co | 271.7 d | 122 | $6\times10^4$ | $10^{-3}$ | $10^{-3}$ |
| $^{63}$Cu | 69.15 | (n,2n) | 11.0 | $^{62}$Cu | 9.67 min. | 511 | $4\times10^4$ | 47.8 | 11.4 |

**Table II.** List of candidates for residual and neutron activation measurements. The columns show, in order, the initial isotope, its abundance, the reaction type, the reaction threshold, the residual isotope, its half-life and the gamma-decay energy detected via spectroscopy. The final three columns show the estimated total number of residuals, the activity at time zero, and the activity 20 minutes after the shot. These estimates come from the Monte-Carlo simulations detailed in Section III. Note that, as detailed in the main text, the estimates of the neutron-activation measurements are made considering a 25 mm diameter sample, of 10 mm thickness, placed 40 mm from the target. This is not the exact experimental setup, but it aided in designing the activation stack used in the experiment.

Figure 8 shows the cross-sections of the materials considered for this neutron activation diagnosis as taken from the TENDL library [43]

Based on these estimates, some of the activation materials were excluded from the experiment. For example, both of the reactions in $^{58}$Ni were excluded due to the very low activity, which stems not from the number of activations, but from the very long half-life of the products. The copper sample was excluded, not due to the low activity, but to the fact that the copious gamma-rays produced in the laser interaction may create activity in the sample via the $^{63}$Cu(γ,n)$^{62}$Cu or $^{65}$Cu(γ,n)$^{64}$Cu reactions, which produce products indistinguishable from the neutron-generated reactions: $^{63}$Cu(n,2n)$^{62}$Cu and $^{65}$Cu(n,2n)$^{64}$Cu, respectively.

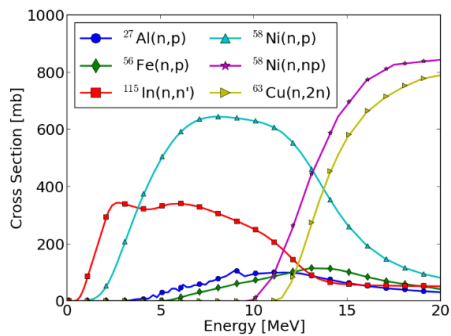

**Fig. 8.** Cross-sections for neutron induced reactions of interest for activation analysis.

The materials that were chosen for this experiment were thus $^{27}$Al, $^{56}$Fe and $^{115}$In, which were used in their natural abundances. Figure 1 shows the activation pack, which consisted of, in order of proximity to the LiF neutron-source, an Indium cylindrical puck (thickness = 3mm, diameter = 25 mm, ρ = 7.31 g/cc), a steel slab, which we approximate as Iron (thickness = 12.5 mm, area = 20x30 mm$^2$, ρ = 7.87 g/cc), an Aluminum cylindrical puck (thickness = 10 mm, diameter = 25 mm, ρ = 2.69 g/cc), and followed by the CR-39, which will be explained in Section IV C.

This stack was placed at 0° from the direction of the proton propagation and at either 45 or 205 mm from the front of the Indium to the LiF. The solid angle, Ω, was calculated as a function of the angle θ subtended by the activation sample with radius R and distance D from the source.

$$\Omega = 4\pi \sin(\frac{\theta}{2})^2 = \frac{\pi R^2}{R^2 + D^2} \qquad (3)$$

In the case of the square shaped Iron, an effective radius of $R^2 = S/\pi$ was used, where S is the surface of the detector piece. The results from these measurements are shown in Table III.

| | | Distance [mm] | A(0) [Bq] | N [n] | dN/dΩ [n/sr] |
|---|---|---|---|---|---|
| Shot 25 | Al | 220.5 | 0 | | |
| | Fe | 208 | 0 | | |
| | In | 205 | 0.8 | $1.9\times10^4$ | $1.6\times10^6$ |
| Shot 42 | Al | 60.5 | 45 | $3.7\times10^4$ | $2.9\times10^5$ |
| | Fe | 48 | 3.42 | $4.6\times10^4$ | $1.9\times10^5$ |
| | In | 45 | 7.52 | $1.8\times10^5$ | $7.8\times10^5$ |

**Table III.** Measured activities at shot time, A(0), the corresponding number of nuclei, N, and the number of nuclei per solid angle, dN/dΩ.

### B. Neutron Time-of-flight diagnostics
#### 1. Simulation of the nToF response

To analyze the experimental nToF signals, simulations were performed using the Geant4 toolkit, as it allowed for modeling the scintillation processes in the detectors. The simulated neutron energy distributions of Figure 7 were used as an input for Geant4. The experimental setup and the interactions of the neutrons in the experimental hall were simulated and nToF temporal distributions were obtained.

The simulations included the Al interaction chamber, the concrete walls of the experimental hall with a thickness of 40 cm and the Pb configurations shielding of each of the detectors. The response of the BC-400 scintillators in terms of scintillation photon yield was set according to Refs. [44,45] and the neutron interactions were simulated using the QGSP_BIC_HP physics list. The simulated nToF distributions were convolved with a Gaussian with a Full Width at Half Maximum (FWHM) of 10 ns that is the measured detector response to a single interaction.

The effects of the different components of the experimental setup on the nToF signals were studied through simulations. Figure 9 presents the simulated nToF signals, under different configurations, for Detector #10 (placed at 191° and 5.14 m from the target, see Figure 2). The distributions in the figure are normalized to each other according to the total number of events registered with the detector.

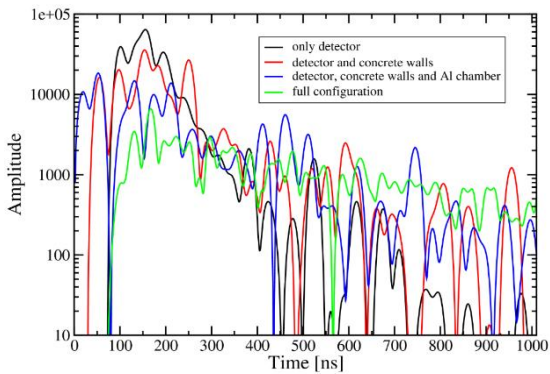

**Fig. 9.** Effects of the addition of different volumes in the simulations of the nToF signals for Detector #10.

The changes in the shape of the nToF signals are clearly observed as more of the environmental surroundings are added to the simulated configuration. The scattering effect can also be visualized in a two-dimensional graph where the relation between the time of flight for the detected event and the initial neutron energy are plotted. Such a plot for Detector #10, in the full configuration simulation is presented in Figure 10. The points correspond to events in the detector and it is clearly visible that wide ranges of time of flight values correspond to neutrons with the same initial energy.

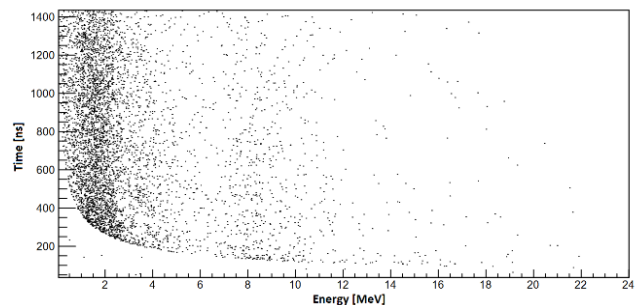

**Fig. 10.** Relationship between the simulated time of flight distribution and the initial energy of the neutrons, for Detector #10, with a 30 cm thick front Pb shielding.

## 2. nToF measurements

The response of the detector was the result of neutron and X/gamma-ray interactions with the scintillator. However, since the X/gamma-rays travel at the speed of light, whereas the neutrons are much slower, there is a temporal separation between the two induced signals, as can be seen in Figure 11. A shielding comprising several Pb bricks was placed around each detector to reduce the saturation effects due to the strong hard X-ray flash associated with the laser interaction with the (primary) target [23,46]. The thickness of the Pb shielding in the direction of the Al interaction chamber was in the range of 20 cm to 40 cm for the different detectors. The arrangement of the scintillators and their shielding is presented in Figure 2.

The neutron time of flight was measured relative to an external trigger connected with the laser pulse. The signals of the detectors were aligned in time using the γ-peak position in each of the spectra. A background subtraction procedure was performed using data from laser shots with similar experimental conditions to the shot of interest, but with the secondary LiF target removed. The background distributions for each detector were fitted with multicomponent exponential functions from the measured nToF spectrum of the corresponding detector. This procedure is illustrated in Figure 11 for Detector #10, with a 30 cm thick front Pb shielding.

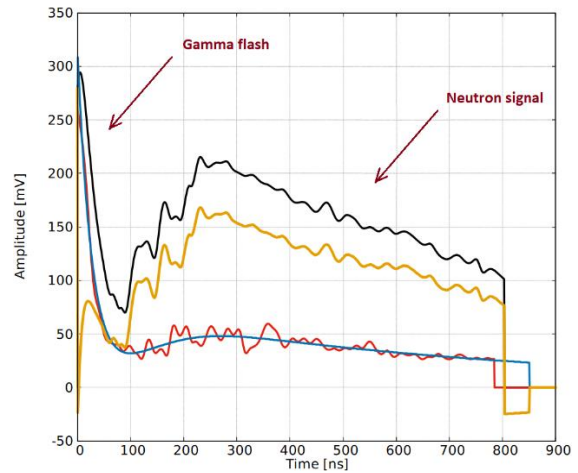

**Fig. 11.** Background subtraction procedure in the experiment for Detector #10, with a 30 cm thick front Pb shielding: (black) nToF for the detector in the shot of interest after the γ-peak time alignment; (red) background distribution from a shot with the LiF secondary target removed; (blue) fit of the background distribution; (yellow) background subtracted nToF distribution for the detector.

### C. CR-39 measurements

Another yield diagnostic used in this experiment was CR-39, which is a plastic that is damaged by knock-on (i.e. elastically scattered) protons and other ions that are scattered by the incoming neutrons [47]. These ions damage the plastic and are then revealed as tracks once the CR-39 has been etched in a strong base, which in our case was 6.0 molarity NaOH at 80°C for 6 hours.

As shown by Frenje, et al. [47], the efficiency of the track production depends not only on the etching parameters, but also on the side of the plastic that is etched. This is because neutron-scattered ions will generally have a forward-focused trajectory, which decreases the number of tracks on the front (facing the neutron source) of the CR-39. However, to avoid this issue and increase the number of tracks observed, the CR-39 was used only in stacks of 4 pieces, where the first layer was re-used and not counted. Thus, because all of the pieces were bordered by CR-39, there should be no differential between front and rear, allowing all pieces to be treated as one would treat the rear side. The efficiency of the pit production, for these etching parameters and for the "rear" side, is shown in Figure 12.

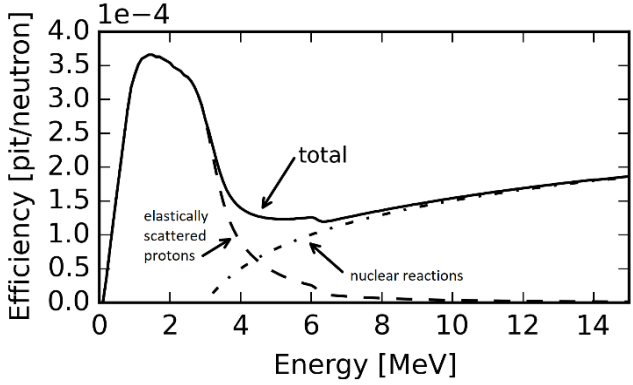

**Fig. 12.** The efficiency of CR-39 when etched in 6.0 molarity NaOH at 80°C from 6 hours. These numbers correspond to the "rear-side of the plastic" as detailed in the text.

To count the tracks in the CR-39 after etching, the pieces were viewed in a microscope with a zoom that covered 0.57 mm² and 25 pictures were taken for a total area of 14.52 mm². As noted, a total of 3 pieces were used for each location. An un-irradiated sample of CR-39 was also etched and counted, which showed a total of 27 tracks over the observed sample (1.9 tracks/mm²). The resulting data from the CR-39 study is shown in Table IV, where the total of all three pieces of CR-39 is shown in the "Meas." column after subtracting the background from the unexposed piece. The number of tracks per steradian is calculated by the same procedure as in equation (3), using the same method to calculate an effective radius as with the aluminum.

|  | Angle [deg] | Distance [mm] | Meas. [track] | dMeas./dΩ [track/sr] |
|---|---|---|---|---|
| Shot 25 | 0 | 232 | 214 | $2.7\times10^5$ |
|  | 18 | 232 | 86 | $1.1\times10^5$ |
|  | 90 | 277 | 12 | $2.1\times10^4$ |
|  | 138 | 151 | 88 | $5.9\times10^4$ |
| Shot 42 | 0 | 45 | $3.7\times10^4$ | $3.9\times10^4$ |
|  | 18 |  |  |  |
|  | 90 | 7.52 | $1.8\times10^5$ | $1.3\times10^5$ |
|  | 138 |  |  |  |

**Table IV.** Experimental results from the CR-39 measurements. The tracks are counted over 25 photographs of 0.57 mm² area on 3 pieces stacked behind each other. The sum of the number of tracks for all three pieces is shown in the third column. The fourth column uses the average of the three pieces using the solid angle calculated from equation (3).

## V. COMPARISON BETWEEN EXPECTED AND MEASURED NEUTRONS

### A. Expected yields using simulated neutron spectra

We will here now describe how the neutron yield measurements are simulated, and also the spectral sensitivity of the activation measurements, due to the energy-dependence of the cross-sections. To aid in this study, we define a few variables and, for ease of description, we describe both the CR-39 tracks and the activated residuals as "events".

$$\frac{dN}{dEd\Omega}(E) = \text{neutron spectrum, [n/MeV/sr]}$$

$$\sigma(E) = \text{cross-section, [m}^2\text{]}$$

$$n_i = \text{number density, [m}^{-3}\text{]}$$

$$\chi = \text{isotope abundance}$$

$$\Omega = \text{solid angle, [sr]}$$

$$t = \text{thickness of sample, [m]}$$

$$\eta(E) = \text{efficiency, [events/neutron]}$$

$$\frac{da}{dE}(E) = \text{activity spectrum, [event/MeV]}$$

$$A = \text{number of nuclei, [events]}$$

$$p(E) = \frac{1}{A}\frac{da}{dE} = \text{activity prob. density function, [1/MeV]}$$

For the CR-39, the efficiency is shown in Figure 12. The efficiency of the activation samples was derived as follows:

$$\eta(E) = \sigma(E) \times n_i \times t \times \chi \quad (4)$$

The activation spectrum, $\frac{da}{dE}(E)$, was obtained by multiplying the neutron spectra arriving on the sample (calculated in Section III) by the efficiency of the reaction. This value gives an indication of which neutrons are primarily responsible for creating events in the samples.

$$\frac{da}{dE}(E)_{sim} = \Omega \times \eta(E) \times \frac{dN}{dEd\Omega}(E)_{sim} \quad (5)$$

$$A_{sim} = \int \frac{da}{dE}dE_{sim} \quad (6)$$

The activation spectra of the samples are plotted in Figure 13, which shows that each sample is dominated by neutrons within a certain energy range.

To define the average energy and neutron number calculated from the simulations, we multiply these quantities by the activity probability density function, $p(E)$, and integrate over $E$.

$$\langle E \rangle_{sim} = \int E \times p(E) dE \quad (7)$$

$$\left\langle \frac{dN}{dEd\Omega} \right\rangle_{sim} = \int \frac{dN}{dEd\Omega}(E) \times p(E) dE \quad (8)$$

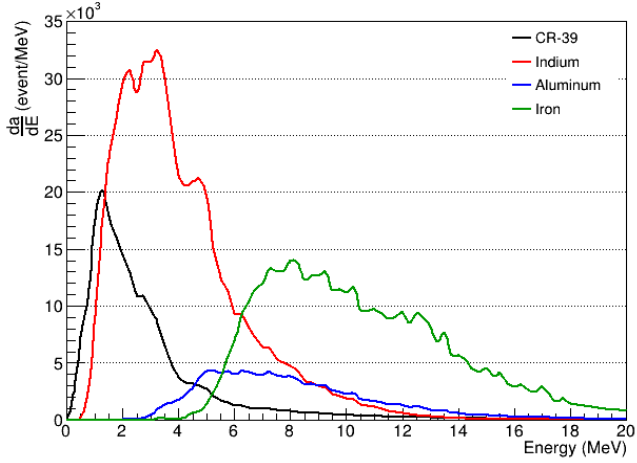

**Fig. 13.** The activation spectrum, $\frac{da}{dE}(E)_{sim}$, of the neutron yield measurements used in the experiment.

Then, the simulated neutron fluences are normalized by the measured number of nuclei, using equation (9), to obtain experimental neutron fluences at the average energies previously calculated.

$$\left\langle \frac{dN}{dEd\Omega} \right\rangle_{exp} = A_{exp} \times \frac{\left\langle \frac{dN}{dEd\Omega} \right\rangle_{sim}}{A_{sim}} \quad (9)$$

The simulated average energies, simulated and experimental neutron fluences are shown in Table V.

| | Angle [deg] | $\langle E \rangle$ [MeV] | $\left\langle \frac{dN}{dEd\Omega} \right\rangle_{sim}$ [n/MeV/sr] | $\left\langle \frac{dN}{dEd\Omega} \right\rangle_{exp}$ [n/MeV/sr] | |
|---|---|---|---|---|---|
| | | | | Shot-25 | Shot-42 |
| CR-39 | 0 | 2.80 | $4.0 \times 10^7$ | $1.9 \times 10^8$ | $2.8 \times 10^7$ |
| | 90 | 2.38 | $2.5 \times 10^7$ | $1.7 \times 10^7$ | $1.1 \times 10^8$ |
| | 180 | 2.02 | $2.9 \times 10^7$ | $4.7 \times 10^7$ | |
| In-115 | 0 | 4.07 | $3.1 \times 10^7$ | $7.9 \times 10^7$ | $3.8 \times 10^7$ |
| Al-27 | 0 | 8.01 | $1.0 \times 10^7$ | | $1.3 \times 10^7$ |
| Fe-56 | 0 | 10.21 | $4.4 \times 10^6$ | | $1.9 \times 10^7$ |

**Table V.** Average energies and neutron fluences for different materials and at different angles.

The average neutron energy examined by each CR-39 foil decreases with increasing angle, which is due to the kinematics of the reaction. This decrease in neutron energy increases the average efficiency, because the CR-39 efficiency peaks at lower energy. The change in average neutron energy with angle is about 20%, which provides an indication of the relative error that would arise from using a single number for the CR-39 efficiency, which has often been done in the past. Also, we would expect this effect to increase with a neutron source that is less isotropic in energy-spectrum (e.g. low-Z (d,n) reactions).

As expected, we probe a range of average energies by using different activation samples. These numbers in Table V correspond to the averaged and total numbers shown in the activation spectrum in Figure 13.

### B. Comparison of expected and measured yields

To compare the simulated neutron yield to the measurements, we use the simulated and experimental neutron fluences calculated at the average energies.

Figure 14 shows the simulated neutron spectrum as a function of energy. The line shows the Geant4 neutron spectrum calculated in Section III. The markers correspond to the simulated and measured data, using the average efficiency and energies. The data points from Shot 42 are close to the simulated values, especially for the CR-39 and the indium sample. However, on the other hand, the data from Shot 25 are quite a bit higher than the expected results.

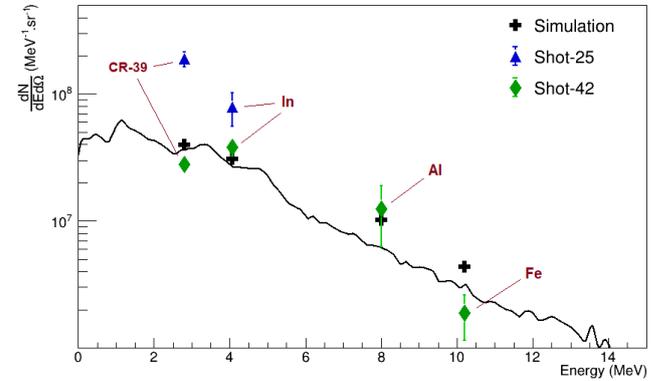

**Fig. 14.** Neutron energy spectrum at 0°. The dark line shows the spectrum determined from the Geant4 simulations which has been normalized by the number of neutrons observed from the $^7$Be residual (i.e. 2 times). The different markers represent the average energies and neutron fluences for individual samples as described in the text.

Figure 15 shows the neutron number as a function of angle. All of the data points are from the CR-39, which has an average energy of around 2 MeV. We find that the experimental data points straddle the expectations from simulations.

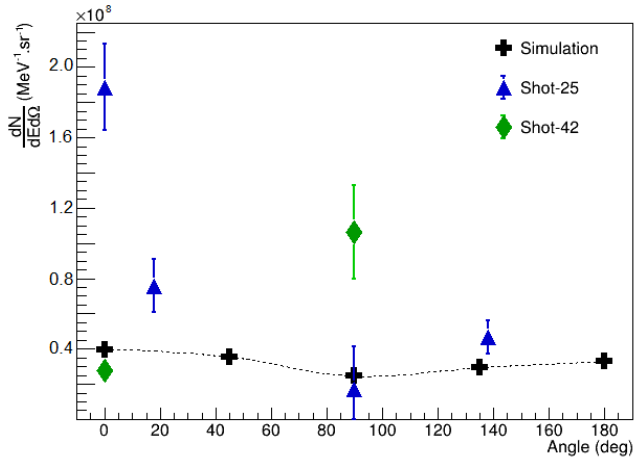

**Fig. 15.** Expected and measured neutron fluence of the CR-39 detectors at different angles around the target. The fluence corresponds to the average fluence in the CR-39, which has an average energy of around 2 MeV. The cross symbols with a dashed line show the simulated data, which has been normalized by the $^7$Be residual activity measurement (i.e. multiplied by 2) and the other markers are from the measurements.

### C. Comparison between expected and nToF measurements

A comparison between the simulated and background subtracted experimental nToF distributions for a laser shot with an energy of 114.8 J is presented in Figure 16. The four plots show distributions for detectors at similar angles but different distances. A correction was applied such that all of the detectors would be placed at a fixed distance of 2 m away from the LiF converter. The thin and thick lines of the same color represent respectively the simulated and experimental nToF for a particular detector.

The Geant4 simulated distributions were obtained in terms of scintillation photon number as function of nToF. For comparison with experimental data they were converted to voltage signals using results from a calibration that was separately performed using a $^{60}$Co and $^{137}$Cs gamma sources to find the Compton edge of the relevant gamma energy. The simulated scintillation photon yield (mainly due to Compton electrons) was calibrated against the electric signal charge, i.e. time integral of current intensity corresponding to the measured voltage on a 50 Ω load. In such a way, the scaling of the height of the simulated nToF distribution to the experimental spectra can be used to estimate the number of neutrons produced in a laser shot.

We have shown for the laser conditions described in this experiment, the parts of the experimental and simulated distributions at high nToF are in a good agreement. Significant discrepancies are observed at low nToF in scintillators close to the interaction chamber. The experimental and simulated distributions fit well for detectors at larger distances. A possible explanation of this effect is an alteration of the signal due to saturation effects in the photomultiplier due to the intense neutron flux for the detectors closer to the secondary target.

An estimation of the neutron flux was obtained on a comparison between the simulated and experimental nToF distributions for Detector #10, with a 30 cm thick front Pb shielding. The results suggest a value of ∼2.0×10$^9$ neutrons in the laser shot, which is in a good agreement with the values obtained by the other methods detailed in this work.

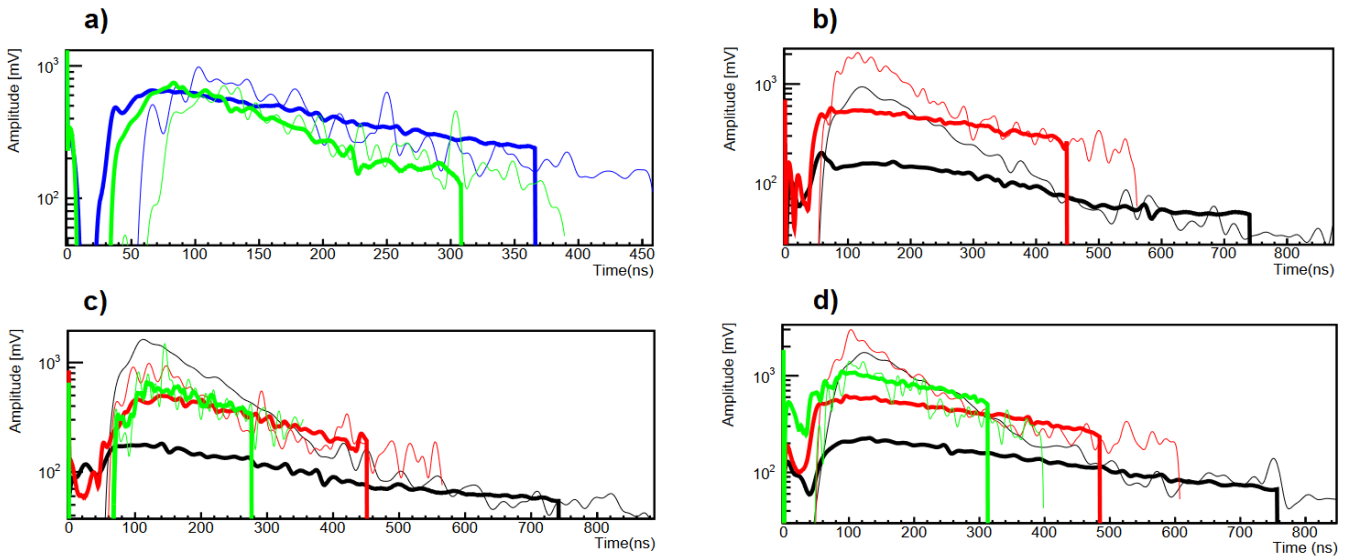

**Fig. 16.** Comparison between simulated and experimental nToF distributions. The four plots represent groups of detectors at similar angles but different distances from the secondary LiF target with (a) Detector #1 and #2 placed at 34 degrees, (b) Detector #3 and #4 at 124 degrees, (c) Detector #5, #6 and #7 at 152 degrees and (d) Detector #8, #9 and #10 at 191 degrees. The thin and thick lines of the same color represent respectively the simulated and experimental nTOF signals for a particular detector. A correction in the nToF distributions was applied such that all of the detectors would be like placed at a fixed distance of 2 m away from the secondary target.

## V. CONCLUSION

In this paper, we have characterized the yield and angular distribution of laser-produced neutrons. These neutrons were produced from the interaction of protons, generated via the TNSA mechanism using the Titan laser, and LiF converters. We used different independently operating diagnostics, to make comparisons between the different measurements carried out, and thus have more confidence in the results obtained.

The neutron source characterization was done using several diagnostics, including CR-39, activation samples, Time-of-Flight detectors and direct measurement of $^7$Be residual nuclei in the LiF converters. Monte-Carlo simulations were used to simulate the energy spectrum and angular distribution of the neutrons and compared against the experimental data.

The measurements of the LiF converters using gamma spectrometry revealed an average neutron production of about $2.0 \times 10^9$ neutrons per shot, approximately 2 times greater than the simulated value.

A deconvolution procedure, using the simulated neutron energy distribution, was followed to obtain the average energy at which the CR-39 and each of the activation samples are sensitive, and their respective efficiency. A comparison between the simulated neutron spectrum, scaled by the number of neutrons observed in the measurement of the $^7$Be residual activity, and the results obtained by these passive diagnostics showed fairly small differences between the expected and measured values, especially for the activation samples.

The simulated neutron spectra at different angles were also used in conjunction with the modeling of the Time-of-Flight detectors scintillation processes to simulate the expected nToF signals for detectors placed at different angles and distances from the target chamber center. A scaling of these expected nToF signals was made to fit the measured ones. This led to good agreement between simulated and measured signals when a total number of $2.0 \times 10^9$ neutrons was considered in the simulation. Only a few discrepancies, caused by saturation effects on the detectors placed closest to the chamber, were observed for the earliest part of the nToF signals. The good agreements obtained for detectors placed at different angles confirm the expected results regarding the neutron angular distribution.

## VI. ACKNOWLEDGMENTS

The authors thank S. Andrews, J. Bonlie, C. Bruns, R. C. Cauble, D. Cloyne, R. Costa and the entire staff of the Titan Laser and the Jupiter Laser facility for their support during the experimental preparation and execution. This work was supported by funding from the European Research Council (ERC) under the European Unions Horizon 2020 research and innovation program (Grant Agreement No. 787539, Project GENESIS), by Grant No. ANR-17-CE30-0026-Pinnacle from the Agence Nationale de la Recherche, by CNRS through the MITI interdisciplinary programs, by IRSN through its exploratory research program, and by the UC Laboratory Fees Research Program. This work was performed under the auspices of the U.S. Department of Energy by LLNL under Contract DE-AC52-07NA27344 and by Lawrence Berkeley National Laboratory under Contract DE-AC02-05CH11231.